\documentclass[slac_one]{revtex4}
\usepackage{graphicx}
\usepackage{wrapfig}

\usepackage{fancyhdr} 
\usepackage{amsmath,amssymb}

\newcommand{\be}{\begin{equation}}
\newcommand{\ee}{\end{equation}}
\newcommand{\bes}{\begin{subequations}}
\newcommand{\ees}{\end{subequations}}
\newcommand{\bea}{\begin{eqnarray}}
\newcommand{\eea}{\end{eqnarray}}
\newcommand{\bear}{\begin{equation}\begin{array}}
\newcommand{\eear}[1]{\end{array}\label{#1}\end{equation}}
\def\ba{$$\begin{array}}
\def\ea{\end{array}$$}
\def\bra{$\begin{array}}
 \def\era{\end{array}$}

\newcommand{\fr}[2]{\dfrac{{ #1}}{{ #2}}}

\newcommand{\la}{\langle}
\newcommand{\ra}{\rangle}

\newcommand{\epe}{\mbox{$e^+e^-\,$}}
\newcommand{\ggam}{\mbox{$\gamma\gamma\,$}}
\newcommand{\egam}{\mbox{$e\gamma \,$}}

\newsavebox{\fmbox}

{\end{list}}
\newcounter{enumct}

\newcommand{\bu}{$\bullet$\ }

\begin{document}

\title{ Charge asymmetries in $e\gamma \to eW^+W^-$.
Hunting for strong interaction in Higgs sector}

\author{ I.~F.~Ginzburg, K.A.~Kanishev}
\affiliation{Sobolev Institute of Mathematics and
Novosibirsk State University, Novosibirsk, 630090, Russia}

\begin{abstract}
The study of charge asymmetry of $W$ bosons in the process
$\egam\to eWW$ can be a tool for discovery of strong
interaction in Higgs sector at  energies that are lower
than it is necessary for observation of resonances caused
by this strong interaction.
\end{abstract}

\maketitle
\section{INTRODUCTION}

It is well known that at large values of Higgs boson
self-coupling constant, the Higgs mechanism of Electroweak
Symmetry Breaking in Standard Model (SM) can be realized
without actual Higgs boson but with strong interaction in
Higgs sector (SIHS) which will manifest itself as a strong
interaction of longitudinal components of $W$ and $Z$
bosons. It is expected that this interaction will be
similar to the interaction of $\pi$-mesons at
$\sqrt{s}\lesssim 1.5$~GeV and will manifest itself in the
form of $W_LW_l$, $W_LZ_L$ and $Z_LZ_L$ resonances. Main
efforts to discover this opportunity are oriented for
observation of such resonance states. It is a difficult
task for the LHC due to high background and it cannot be
realized at the energies reachable at the ILC in its
initial stages.

The experience in low energy physics allows us to suggest
the approach for discovery of SIHS at the second stage of
ILC (with c.m.s. energy $0.8\div 1$~TeV) in the case of
realization of Photon Collider mode \cite{PLC}.

\bu {\bf Some history}.

The study of charge asymmetry of pions in   $\epe\to
\epe\pi^+\pi^-$ allows to measure relative phase of
$\ggam\to \pi\pi$ and $\gamma\to\pi\pi$ amplitude, caused
by strong interaction  \cite{eepipi}. This asymmetry
appears due to interference of amplitudes describing
production of $\pi\pi$ systems with opposite
$C$--parities.\vspace{-10mm}
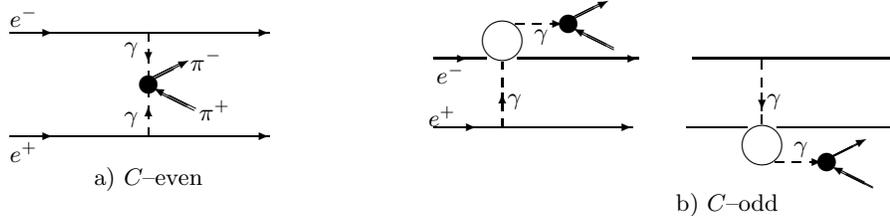
\begin{figure}[hbt]
  \unitlength=1.15mm
    \begin{picture}(42.00,20.00)(10,0)
    \put(19.00,15.00){\makebox(0,0)[cc]{$\gamma$}}
    \put(19.00,7.00){\makebox(0,0)[cc]{$\gamma$}}
    \put(21.00,0.01){\makebox(0,0)[c]{a) $C$--even}}
    \put(35.00,18.80){\makebox(0,0)[r]{$$}}
    \put(5.00,18.80){\makebox(0,0)[l]{$e^-$}}
    \put(35.00,3.20){\makebox(0,0)[r]{$$}}
    \put(5.00,3.20){\makebox(0,0)[l]{$e^+$}}
    \put(21.00,5.00){\line(0,1){1.00}}
    \put(21.00,9.50){\line(0,1){1.00}}
    \put(21.00,17.00){\line(0,-1){1.00}}
    \put(21.00,12.50){\line(0,-1){1.00}}
    \put(21.00,11.00){\circle*{2.00}}
    \put(20.90,11.25){\line(2,1){4.00}}
    \put(20.70,11.45){\line(2,1){4.00}}
    \put(22.50,10.15){\line(2,-1){4.00}}
    \put(22.30,10.00){\line(2,-1){4.00}}
    \put(21.60,10.50){\vector(-2,1){0.00}}
    \put(25.70,13.80){\vector(2,1){0.00}}
    \put(27.70,13.80){\makebox(0,0)[cc]{$\pi^-$}}
    \put(28.70,8.00){\makebox(0,0)[cc]{$\pi^+$}}
    \put(21.00,15.00){\vector(0,-1){1.50}}
    \put(21.00,7.00){\vector(0,1){1.50}}
    \put(05.00,17.00){\vector(1,0){5.00}}
    \put(10.00,17.00){\vector(1,0){25.00}}
\put(05.00,5.00){\vector(1,0){5.00}}
    \put(10.00,5.00){\vector(1,0){25.00}}

  \end{picture}
  \begin{picture}(55.00,30.00)(0,05)
\    \put(21.30,25.70){\vector(2,1){0.00}}
    \put(16.50,23.40){\line(2,1){4.00}}
    \put(16.50,23.20){\line(2,1){4.00}}
    \put(17.20,22.50){\vector(-2,1){0.00}}
    \put(17.90,22.00){\line(2,-1){4.00}}
    \put(17.90,22.20){\line(2,-1){4.00}}
 \put(6.00,11.00){\vector(1,0){18.00}}
    \put(1.00,11.00){\vector(1,0){5.00}}
    \put(10.70,19.00){\vector(1,0){14.00}}
    \put(1.00,19.00){\vector(1,0){4.00}}
    \put(14.50,23.00){\vector(1,0){1.50}}
    \put(44.50,7.00){\vector(1,0){1.50}}
    \put(40.70,19.00){\line(1,0){14.00}}
    \put(31.00,19.00){\line(1,0){4.00}}
    \put(5.00,19.00){\line(1,0){2.20}}
    \put(10.50,23.00){\line(1,0){1.00}}
    \put(12.50,23.00){\line(1,0){1.00}}

\put(55.00,11.00){\line(-1,0){4.00}}
    \put(37.30,11.00){\line(-1,0){7.00}}

    \put(9.00,21.00){\circle{4.60}}

    \put(16.6,23){\circle*{2.0}}

    \put(9.00,13.00){\vector(0,1){2.00}}
    \put(9.00,11.00){\line(0,1){1.00}}
    \put(9.00,15.70){\line(0,1){1.0}}
    \put(9.00,17.80){\line(0,1){0.8}}

    \put(2.00,12.30){\makebox(0,0)[cc]{$e^+$}}
    \put(22.00,12.30){\makebox(0,0)[cc]{$$}}
    \put(10.40,14.00){\makebox(0,0)[cc]{$\gamma$}}
    \put(3.00,17.20){\makebox(0,0)[cc]{$e^-$}}
    \put(22.00,17.20){\makebox(0,0)[cc]{$$}}
    \put(13.50,21.50){\makebox(0,0)[cc]{$\gamma$}}
    \put(35.00,02.00){\makebox(0,0)[cc]{b) $C$--odd}}

    \put(51.00,11.00){\line(-1,0){10.3}}

    \put(51.10,9.70){\vector(2,1){0.00}}
    \put(46.40,7.40){\line(2,1){4.00}}
    \put(46.40,7.20){\line(2,1){4.00}}
    \put(47.10,6.50){\vector(-2,1){0.00}}
    \put(47.80,6.00){\line(2,-1){4.00}}
    \put(47.80,6.20){\line(2,-1){4.00}}
    \put(35.00,19.00){\line(1,0){7.00}}
    \put(40.50,7.00){\line(1,0){1.00}}
    \put(42.50,7.00){\line(1,0){1.00}}
    \put(39.00,9.00){\circle{4.60}}

    \put(39.00,15.00){\vector(0,-1){2.00}}

    \put(46.5,7){\circle*{2.0}}

    \put(39.00,11.40){\line(0,1){1.00}}
    \put(39.00,15.70){\line(0,1){1.0}}
    \put(39.00,17.80){\line(0,1){1.2}}
    \put(40.40,14.00){\makebox(0,0)[cc]{$\gamma$}}
    \put(43.50,8.50){\makebox(0,0)[cc]{$\gamma$}}
  \end{picture}
\caption{The $ee\to ee\pi^+\pi^-$ process amplitudes. Open
circles --- virtual Compton scattering. Black circles ---
amplitudes of subprocesses $\ggam\to \pi^+\pi^-$ (C--even
dipion) and $\gamma\to \pi^+\pi^-$ (C--odd dipion),
modified by strong interaction.}
 \end{figure}

\bu Similarly, the opportunity to discover  strong
interaction in Higgs sector before observation of $W_LW_L$
resonances via the study of the charge asymmetry of $W$ in
$\egam\to eWW$ was proposed in ref.~\cite{SD}. This paper
considers this asymmetry within SM in detail.

\section{The $e\gamma\to eWW$ process}

The SM cross section of the process  is about 10~pb in the
discussed energy interval \cite{GIS} which corresponds with
about $10^6$ expected events per year. Main contribution to
the cross section is given by $e\gamma\to
ee'\gamma^*\otimes \gamma\gamma^*\to WW$ process (like in
fig.~1a).  The charge asymmetry appears due to interference
like in pion case. Therefore, its value grows with increase
of electron transverse momentum. Fig.~2 demonstrates these
facts for the unpolarized photons.

\newpage

\begin{figure}[hbt]
  \includegraphics[width=0.3\textwidth, height=6cm]{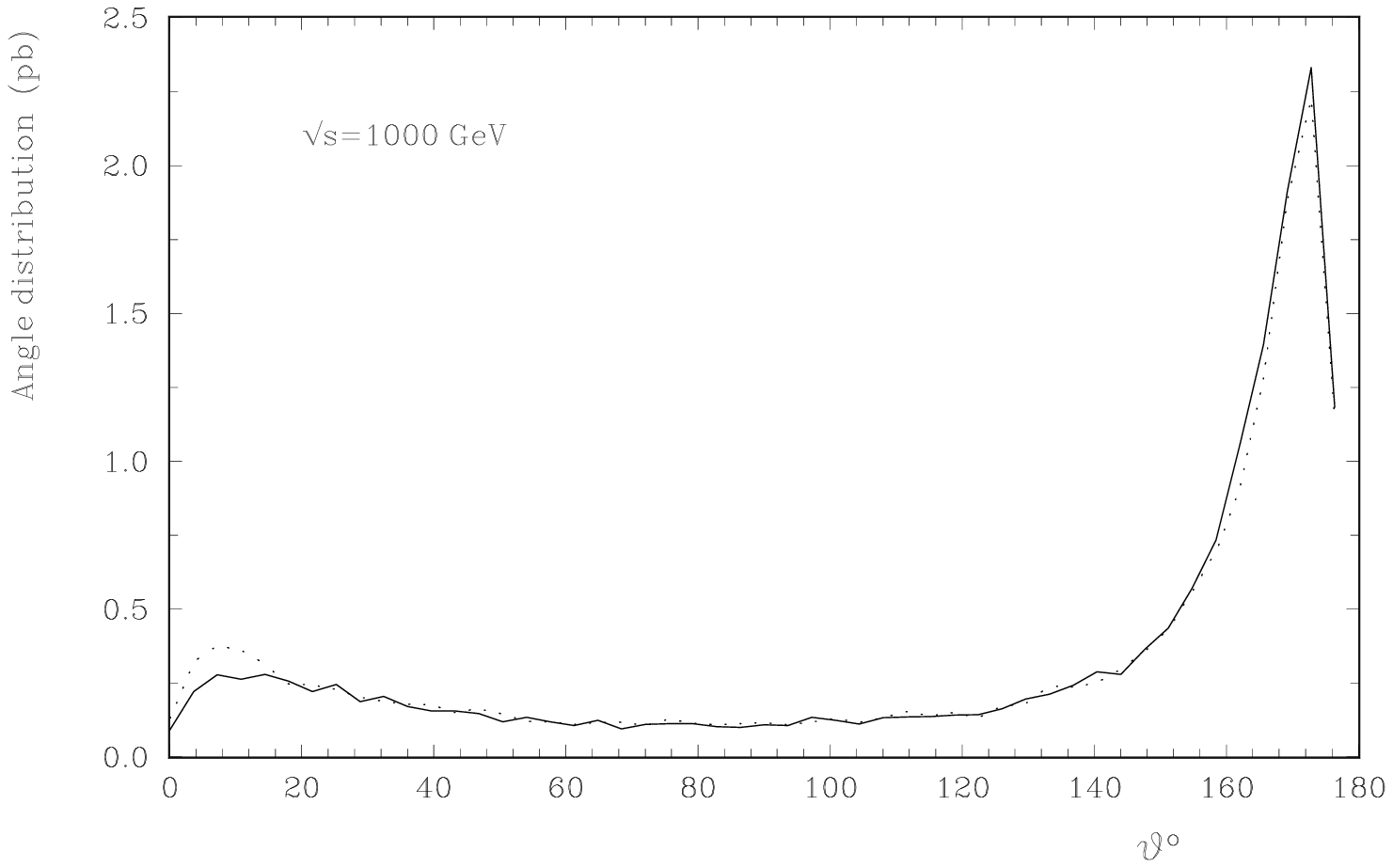}
 \includegraphics[width=0.3\textwidth, height=6cm]{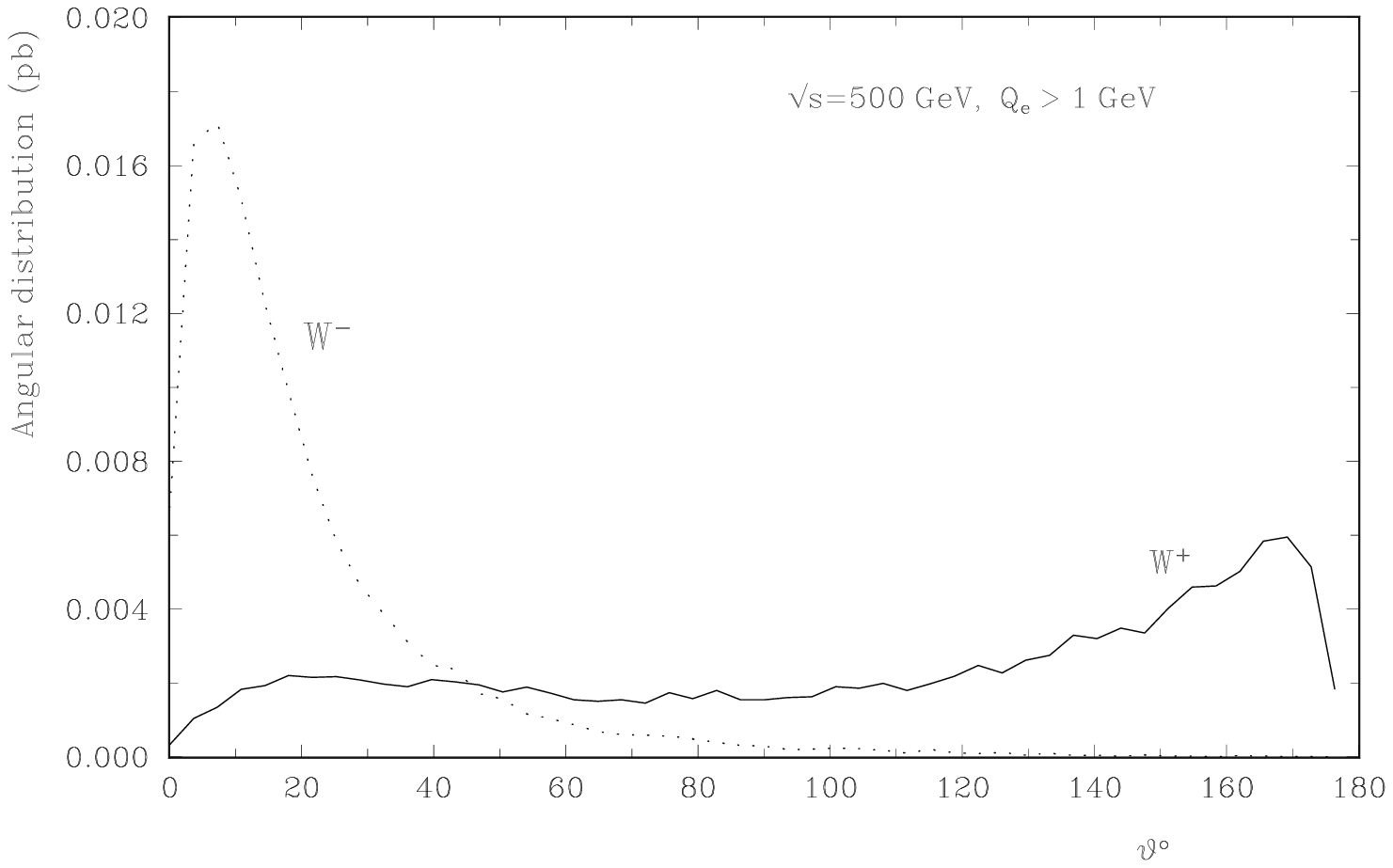}
   \includegraphics[width=0.35\textwidth, height=5cm]{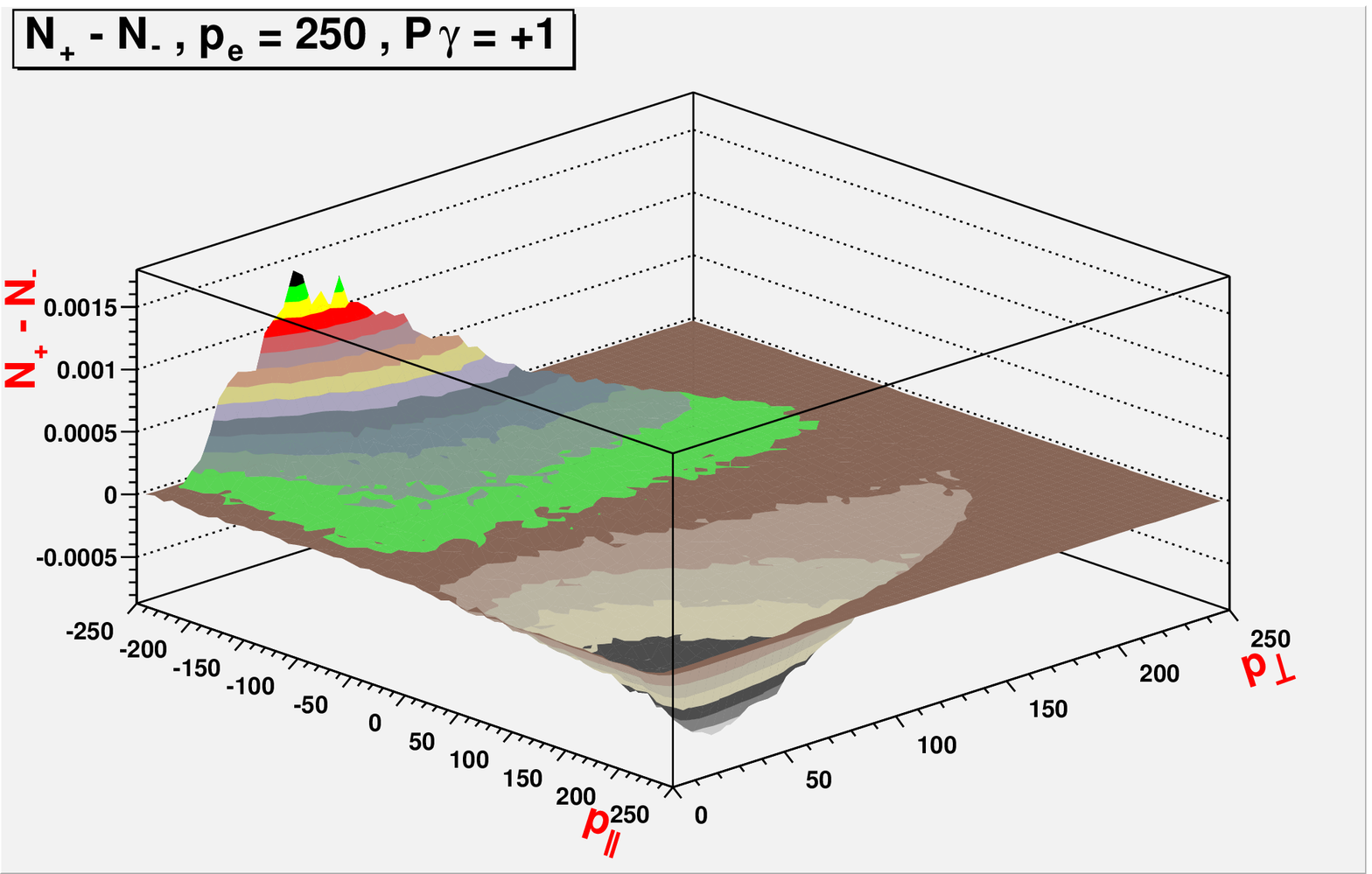}
\caption{Two left graphs (old calculations): Angular
distribution of $W^+$, $W^-$ without cut in $p_\bot$ (left)
and at $p_\bot^e>1$ GeV (center); full -- $W^+$, dotted --
$W^-$, nonpolarized particles \cite{Ilyin}.\\ Right picture
(recent calculations): Difference $N_{W^+}-N_{W^-}$ at
$p_\bot^e>10$~GeV, clockwise circularly polarized photons}
\end{figure}

\subsection{ Diagrams of the process}

The diagrams of the process are subdivided into three types
shown in Fig.~3.
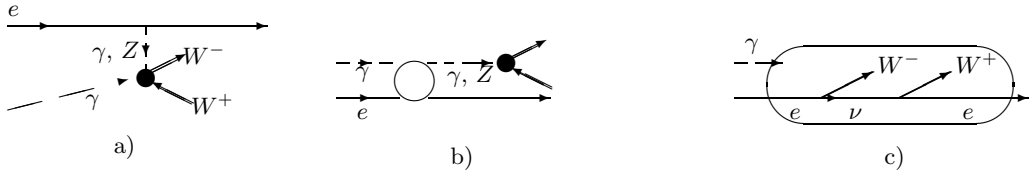
\begin{figure}[hbt]
  \unitlength=1.15mm
    \begin{picture}(132.00,20.00)(10,0)
    \put(17.50,14.00){\makebox(0,0)[cc]{$\gamma,\,Z$}}
    \put(5.00,18.80){\makebox(0,0)[l]{$e$}}
      \put(19.00,9.50){\line(2,1){1.00}}
      \put(8.00,5.00){\line(2,1){1.00}}
 \multiput(5.00,7.32)(6.00,1.5){2}{\line(4,1){4}}
 \put(17,10.32){\vector(4,1){2}}
    \put(14.00,8.2){\makebox(0,0)[l]{$\gamma$}}
    \put(21.00,17.00){\line(0,-1){1.00}}
    \put(21.00,12.50){\line(0,-1){1.00}}
    \put(21.00,11.00){\circle*{2.00}}
    \put(20.90,11.25){\line(2,1){4.00}}
    \put(20.70,11.45){\line(2,1){4.00}}
    \put(22.50,10.15){\line(2,-1){4.00}}
    \put(22.30,10.00){\line(2,-1){4.00}}
    \put(21.60,10.50){\vector(-2,1){0.00}}
    \put(25.70,13.80){\vector(2,1){0.00}}
    \put(27.70,13.80){\makebox(0,0)[cc]{$W^-$}}
    \put(28.70,8.00){\makebox(0,0)[cc]{$W^+$}}
    \put(21.00,15.00){\vector(0,-1){1.50}}
    \put(05.00,17.00){\vector(1,0){5.00}}
    \put(10.00,17.00){\vector(1,0){25.00}}
\put(19.00,3.0){\makebox(0,0)[c]{a) }}
  \end{picture}

  \begin{picture}(65.00,10.00)(0,-10)
  \put(24.30,15.70){\vector(2,1){0.00}}
   \put(19.50,13.40){\line(2,1){4.00}}
   \put(19.50,13.20){\line(2,1){4.00}}
    \put(20.20,12.50){\vector(-2,1){0.00}}
    \put(20.90,12.00){\line(2,-1){4.00}}
    \put(20.90,12.20){\line(2,-1){4.00}}
    \put(10.70,9.00){\vector(1,0){14.00}}
    \put(0.00,9.00){\vector(1,0){4.00}}
\put(4.00,9.00){\line(1,0){1.00}}
    \put(14.50,13.00){\vector(1,0){3.50}}
     \put(0,13){\line(1,0){1.5}}
\put(2.5,13){\vector(1,0){1.5}}
 \put(5,13){\line(1,0){1.5}}
    \put(3.00,11.80){\makebox(0,0)[cc]{$\gamma$}}
    \put(5.00,9.00){\line(1,0){2.20}}
    \put(10.50,13.00){\line(1,0){1.00}}
    \put(12.50,13.00){\line(1,0){1.00}}
    \put(9.00,11.00){\circle{4.60}}
    \put(19.6,13){\circle*{2.0}}
    \put(3.00,7.20){\makebox(0,0)[cc]{$e$}}
        \put(15.50,11.50){\makebox(0,0)[cc]{$\gamma,\,Z$}}
    \put(15.00,2.00){\makebox(0,0)[cc]{b) }}

\put(60.00,9.00){\vector(1,0){20.00}}
    \put(46.00,9.00){\vector(1,0){12.00}}
\put(46,13){\line(1,0){1.5}}

\put(48.5,13) {\vector(1,0){3}}
 \put(48,15){\makebox(0,0)[cc]{$\gamma$}}

\put(56.00,9.00){\line(1,0){4.00}}
 \put(56,9){\vector(2,1){6}}
\put(56.1,9){\line(2,1){6}}
 \put(55.9,9){\line(2,1){6}}
 \put(65,9){\vector(2,1){6}}
 \put(65.1,9){\line(2,1){6}}
  \put(64.9,9){\line(2,1){6}}
    \put(53.00,7.20){\makebox(0,0)[cc]{$e$}}
    \put(60.00,7.20){\makebox(0,0)[cc]{$\nu$}}
      \put(73,7.20){\makebox(0,0)[cc]{$e$}}
\put(65,13){\makebox(0,0)[cc]{$W^-$}}
\put(74,13){\makebox(0,0)[cc]{$W^+$}}
 \put(64,10.5){\oval(28.5,9)}

\put(65.00,2.00){\makebox(0,0)[cc]{c) }}

\end{picture}
\vspace{-15mm} \caption{Types of amplitudes for $e\gamma\to
eWW$ process. }
 \end{figure}

\bu Diagrams of type a) contain subprocesses
$\gamma\gamma\to W^+W^-$ and $\gamma Z\to W^+W^-$, modified
by strong interaction in Higgs sector ({\it two--gauge
contribution}).

\bu Diagrams of type b) contain subprocesses $\gamma\to
W^+W^-$ and $ Z\to W^+W^-$, modified by strong interaction
in Higgs sector  ({\it one--gauge contribution}). Open
circle describes $\egam\to \egam$ or $\egam\to eZ$
subprocesses.

\bu Diagrams of type c) -- with neutrino exchange -- are
built from the diagram shown inside the oval by connecting
the photon line to each charged particle line. Strong
interaction does not modify this contribution.

Diagrams 3a) and 3b) are similar to diagrams 1a) and 1b)
for pions but have an essential difference. Because of $Z$
contribution, corresponding final states have no definite
$C$-parity. Therefore, charge asymmetry appears even within
each this type, like charge asymmetry in \epe\, collision
near $Z$-- peak.

\subsection{Asymmetries in SM}

To see main features of the effect and its potential for
the study of strong interaction in Higgs sector, we
calculated some quantities describing charge asymmetry
(charge asymmetric variables -- CAV) for $e^-\gamma$
collision at $\sqrt{s}=500$~GeV with polarized photons. We
used CompHEP \cite{CompHEP} and CalcHEP \cite{Pukhov}
packages for simulation.

We denoted
 \be
p^\pm \; \mbox{ --  momenta of }\; W^\pm,\qquad
 p_e\; \mbox{ -- momentum of
scattered electron},\quad W=\sqrt{(p^++p^-)^2\,}\,.
 \ee

We studied $W$-dependence of the following averaged
quantities
 \be
 v_1=\fr{\la(p^+-p^-)p_e\ra}{\la(p^++p^-)p_e\ra}\,,\qquad
v_2=
\fr{\la(p^+_\|)^2-(p^-_\|)^2\ra}{\la(p^+_\|)^2+(p^-_\|)^2\ra}\,,\qquad
v_3=
\fr{\la(p^+_\bot)^2-(p^-_\bot)^2\ra}{\la(p^+_\bot)^2+(p^-_\bot)^2\ra}\,.
\ee

We applied the cut in transverse moment of scattered
electron,
 \be
p_\bot^e\ge p_{\bot 0}\;\;\mbox{ with
}\,\left\{\begin{array}{c} a)\; p_{\bot 0}=10~GeV,\\
           b)\; p_{\bot 0}=30~GeV.\end{array}\right.
           \ee
With growth of $p_{\bot 0}$ the two--gauge contribution
decreases strongly while the one--gauge contribution varies
weakly. Therefore, the relative value of the asymmetry
under interest grows. Besides, observation of scattered
electron allows to check kinematics completely.

\bu {\bf Influence of polarization}

First, we consider influence of photon polarization for the
effect. Fig.~4 represents distribution in CAV $v_1$ for
right-hand (upper curves) and left-hand (lower curves)
polarized photons at cuts $p_{\bot 0}=10$~GeV (left) and
$p_{\bot 0}=30$~GeV (right).
\begin{figure}[hbt]
  \includegraphics[width=0.45\textwidth, height=4cm]{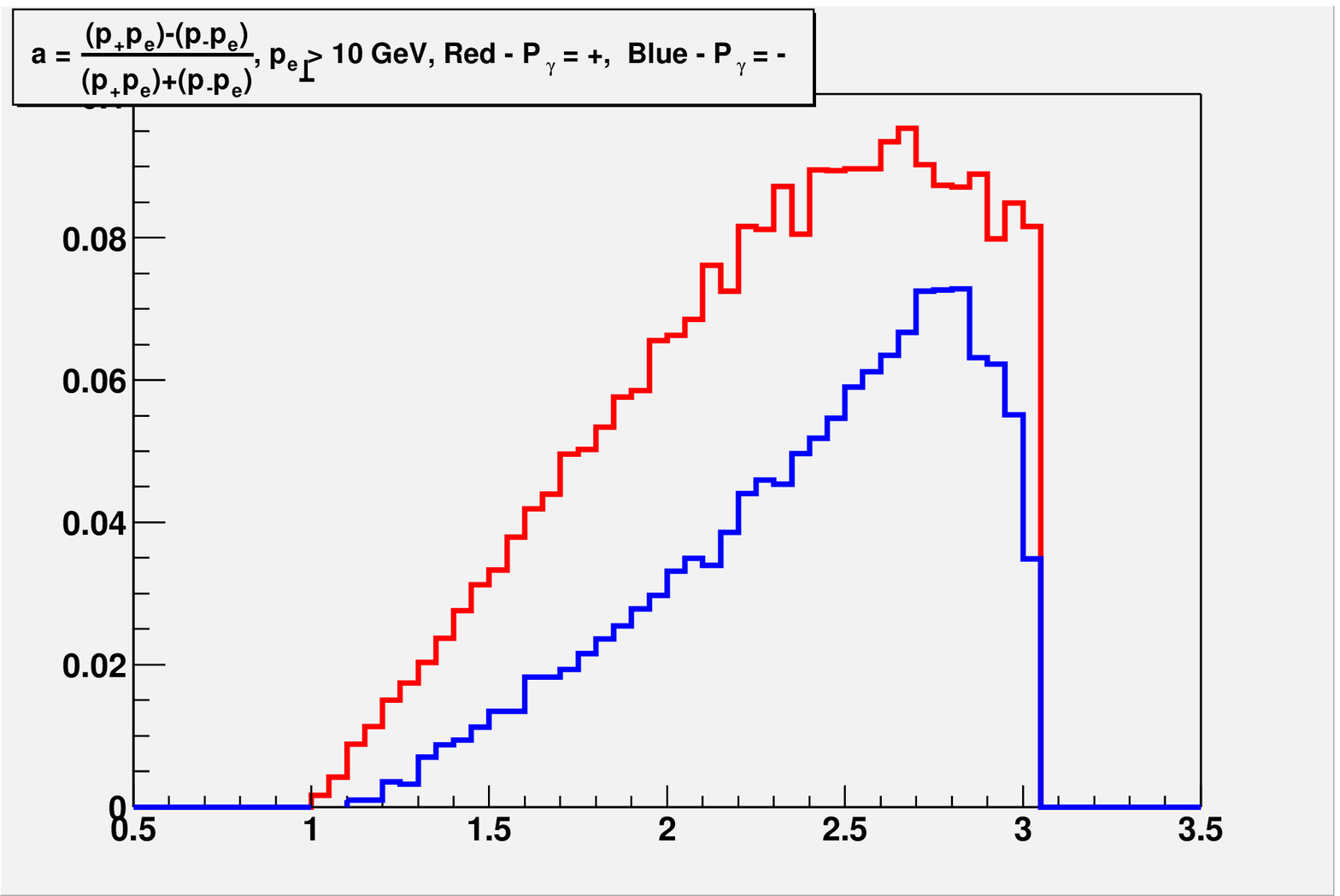}
  \includegraphics[width=0.45\textwidth, height=4cm]{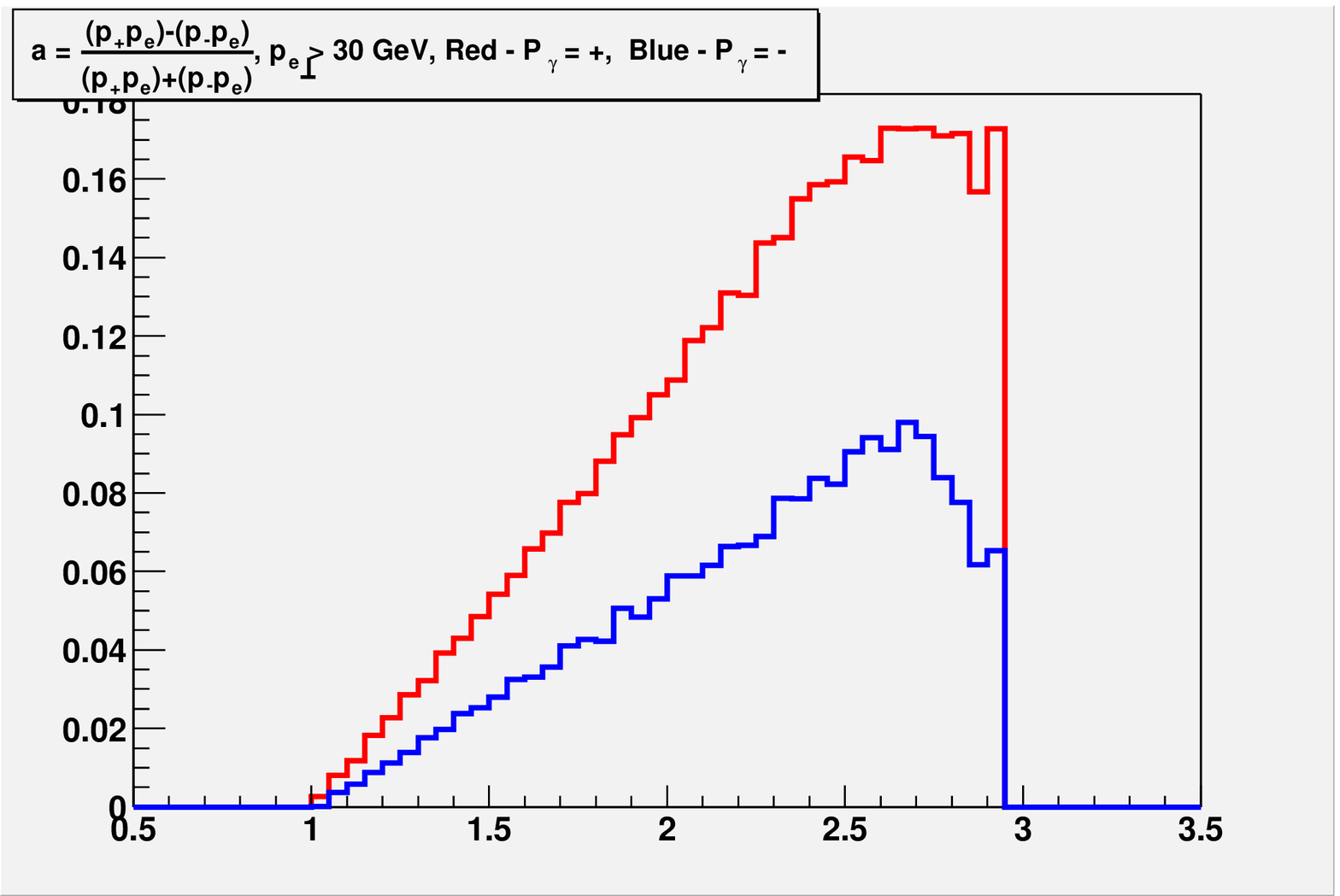}
  \caption{Dependence on polarization and cuts. Variable $v_1$.}
\end{figure}
We do not study the dependence on electron polarization.
This dependence is expected to be weak in SM where main
contribution to cross section is given by diagrams of
Fig.~3a) with virtual photons having the lowest possible
energy. These photons "forget" the polarization of incident
electron. The strong interaction contribution becomes
essential at highest effective masses of $WW$ system with
high energy of virtual photon or $Z$, the helicity of which
reproduces almost completely the helicity of incident
electron \cite{GSerb}. Therefore, study of this
dependence will be necessary part of studies beyond SM.\\

\bu {\bf Significance of different contributions.}

To observe the value of effect under interest, we compared
the entire asymmetry and that without one-gauge
contribution (Fig.~5).
 \begin{figure}
  \includegraphics[width=0.45\textwidth, height=5cm]{pe-+30.eps}
    \includegraphics[width=0.45\textwidth, height=5cm]{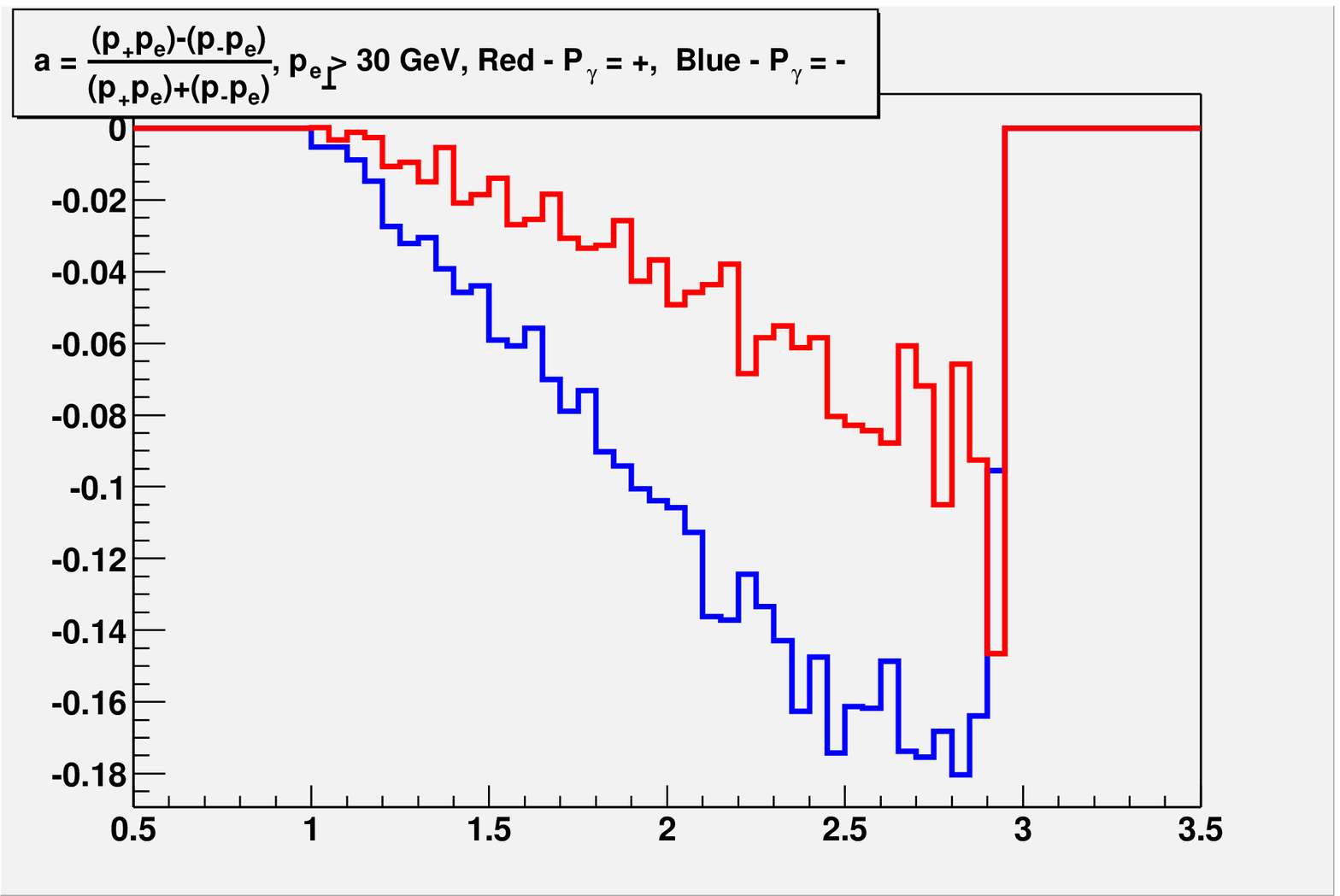}

  \includegraphics[width=0.45\textwidth, height=4cm]{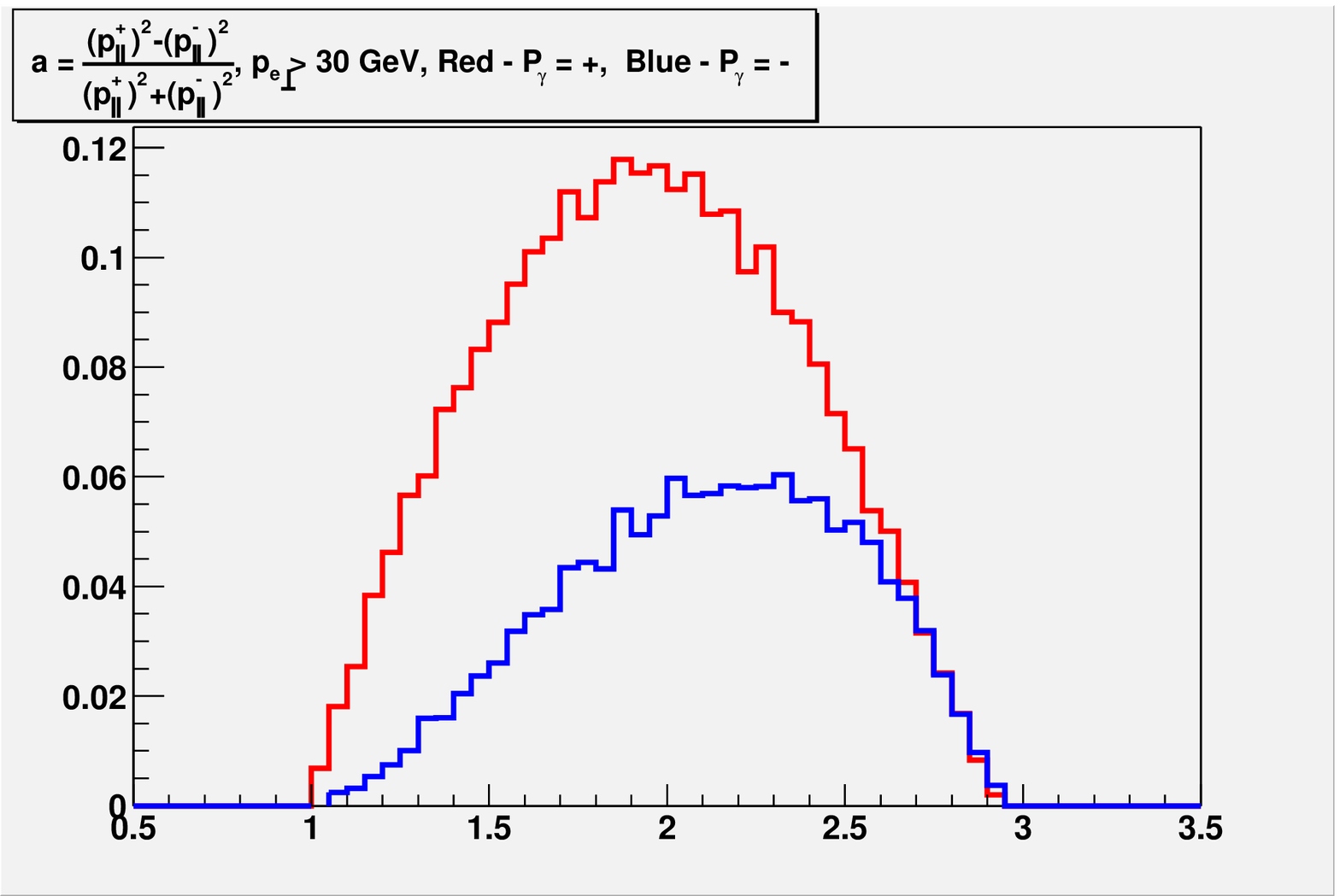}
    \includegraphics[width=0.45\textwidth, height=4cm]{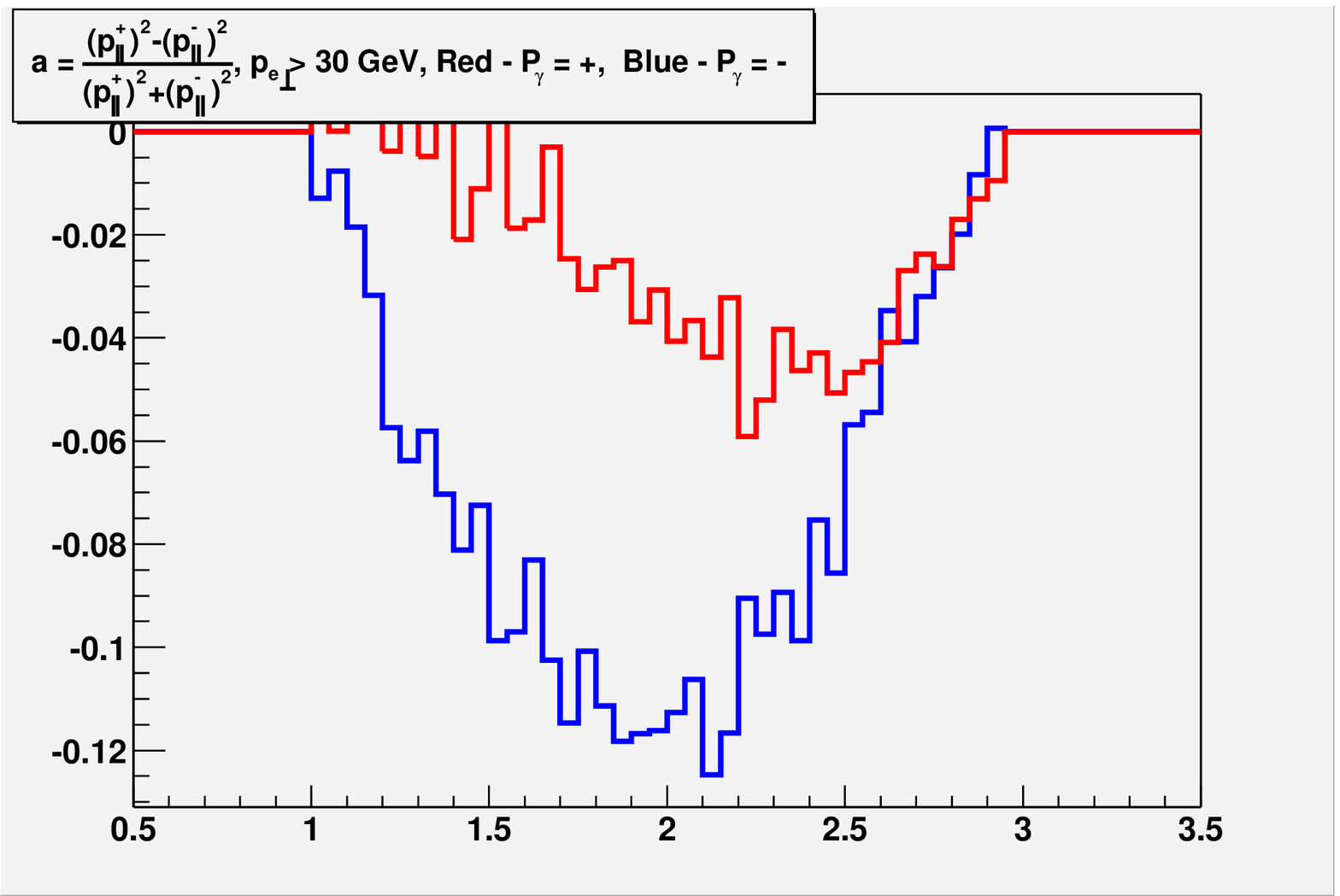}

  \includegraphics[width=0.45\textwidth, height=4cm]{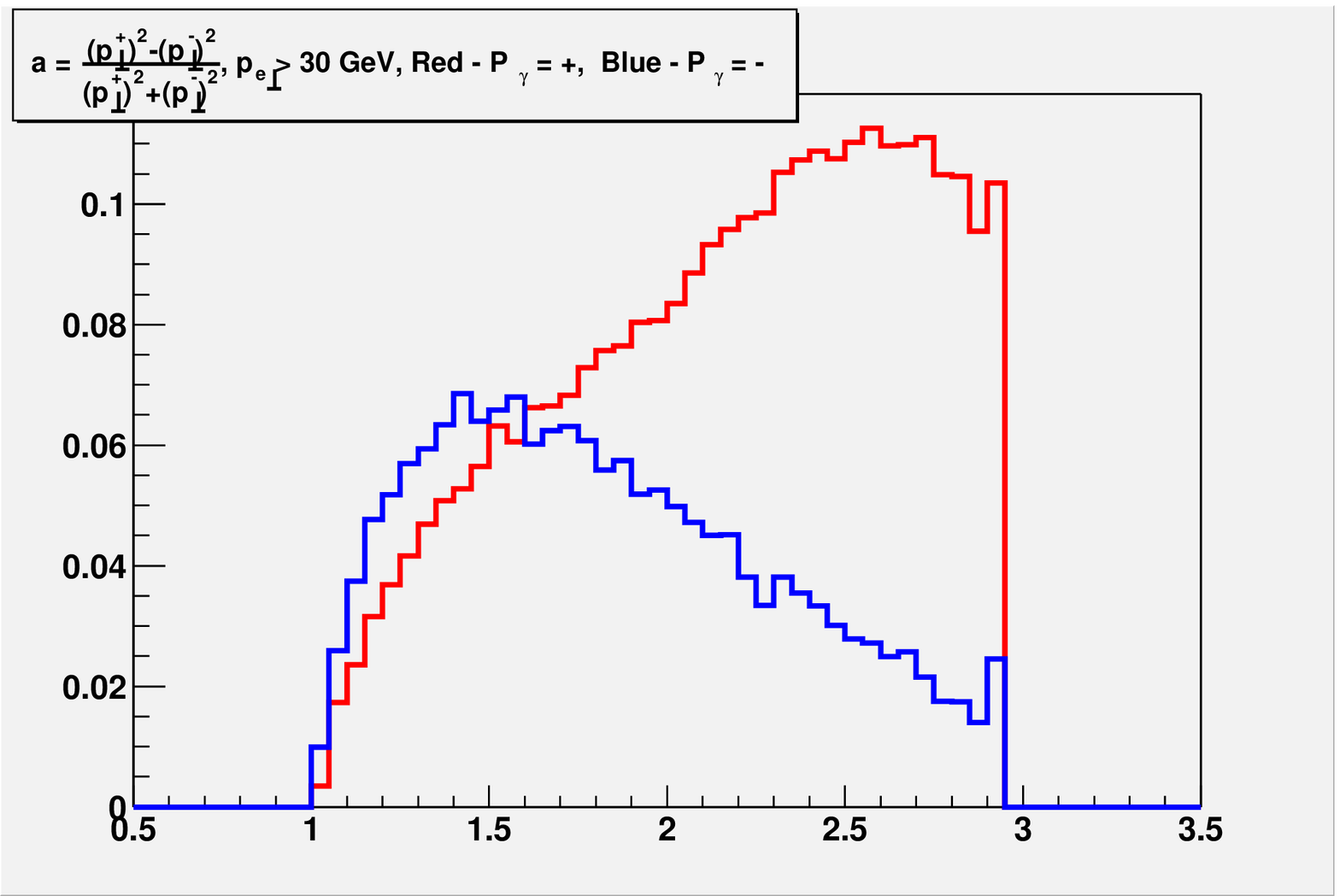}
  \includegraphics[width=0.45\textwidth, height=4cm]{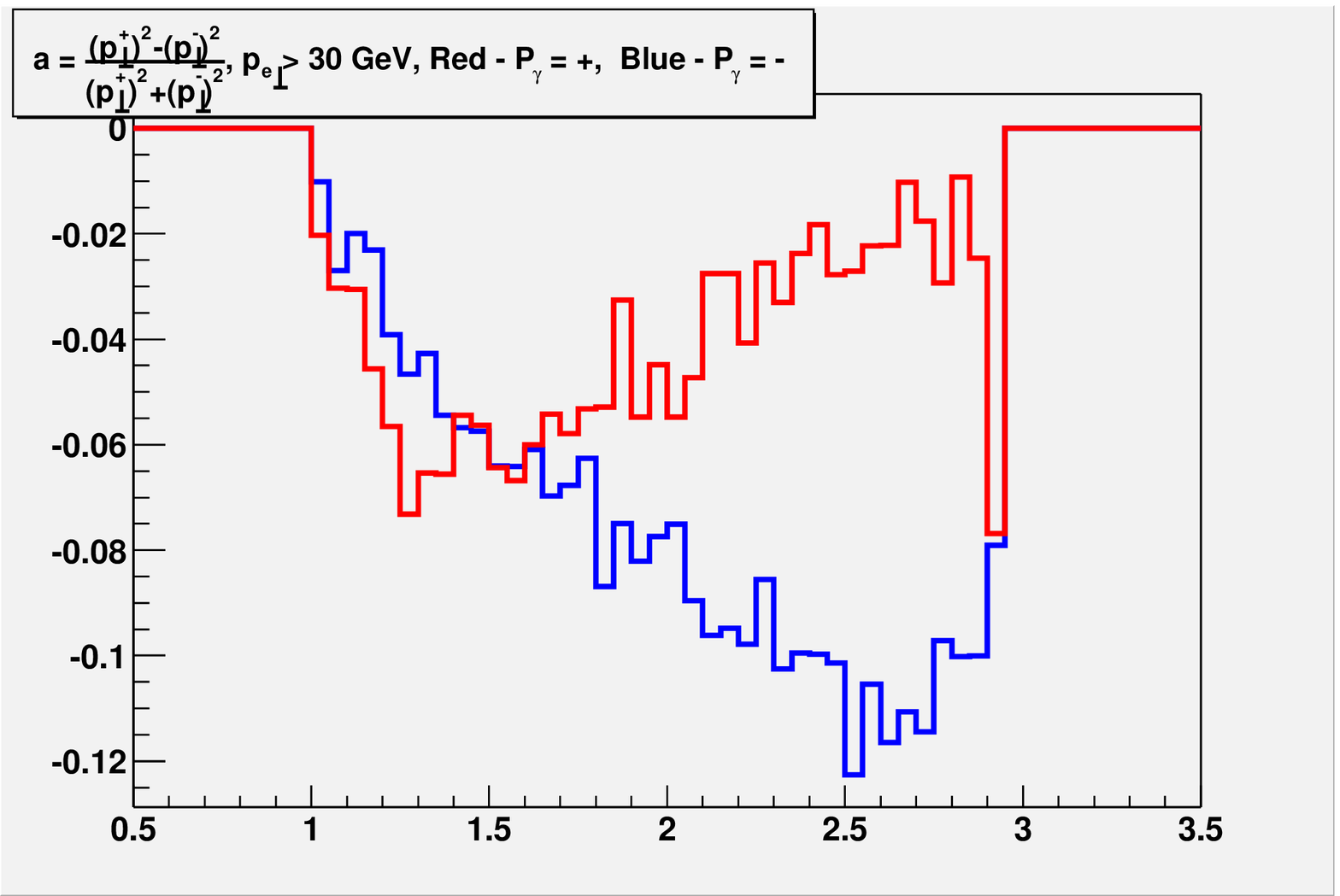}

\caption{Asymmetries in variables $v_1$, $v_2$ and $v_3$
(from top to down). Right -- total, left -- without one
gauge contributions. Upper curves for right--handed
polarized photons, lower  curves for left--handed polarized
photons.}
 \end{figure}
Strong interaction in Higgs sector modifies both one--gauge
and two--gauge amplitudes. The observation of charge
asymmetry caused by their interference will indicate this
strong interaction. One can see that with the account of
one--gauge contribution, the charge asymmetry even changes
its sign.

Therefore, the influence of this potentially informative
contribution to asymmetry is very high.

\newpage
\section{Conclusions and plans}

\bu Charge asymmetry of $W$ is large enough easily
observable effect.

\bu The value of the effect grows with increase of the cut
in $p_{e\bot}$.

\bu  Photon polarization influences strongly the value of
charge asymmetry. The role  of electron polarization
remains to be studied.

\bu  Charge asymmetry is very sensitive to the interference
of two--gauge and one--gauge contributions which modify
under the strong interaction in Higgs sector. Reasonable
approximation for this strong interaction is necessary to
estimate the observable effects. It will be the next step
in our studies.

\begin{acknowledgments}
This work is  supported by grants RFBR 05-02-16211 and
NSh-2339.2003.2. \end{acknowledgments}

\end{document}